\documentclass[floatfix,superscriptaddress,a4paper,
               nofootinbib,preprint]{revtex4}

\usepackage{epsfig}
\usepackage{latexsym}
\usepackage{xspace}

\usepackage{inputenc}
\usepackage{indentfirst}
\usepackage{enumerate}
\usepackage{color}

\usepackage{amsmath}
\usepackage{amssymb}
\usepackage[english]{babel}
\usepackage{url}
\usepackage{hyperref}
\usepackage{xcolor}

\graphicspath{{plots/}}
\include{abbreviations}

\newcommand*{\revision}{\textcolor{black}}

\newcommand{\orcid}[1]{\href{https://orcid.org/#1}{\textcolor[HTML]{A6CE39}{\aiOrcid}}}

\begin{document}

\title{Impact of momentum resolution on factorial moments due to
power-law correlations between particles}

\author{Subhasis Samanta 
}
\email{subhasis.samant@gmail.com}
\affiliation{Jan Kochanowski University, Kielce, Poland}

\author{Tobiasz Czopowicz}
\email{tobiasz.czopowicz@cern.ch }
\affiliation{Jan Kochanowski University, Kielce, Poland}
\affiliation{Warsaw University of Technology, Poland}

\author{Marek Gazdzicki}
\email{marek.gazdzicki@cern.ch}
\affiliation{Jan Kochanowski University, Kielce, Poland}
\affiliation{Geothe-University Frankfurt am Main, Germany}

\begin{abstract}
The effect of momentum resolution on factorial moments due to the power-law correlation function is studied. The study is motivated by searching for the critical point of the strongly interacting matter in heavy-ion collisions using the intermittency method.
We observe that factorial moments are significantly affected by the finite momentum resolution. The effect is superficially significant compared to intuitive expectations.
The results depend on the power of the correlation function and the number of uncorrelated particles. 
\end{abstract}

\pacs{....}
\maketitle

\section{Introduction}
\label{sec:introduction}
The study of the phase diagram of the strongly interacting matter remains one of the main goals of high-energy physics.
Theoretical studies suggest a smooth crossover transition between hadronic and quark-gluon plasma (QGP) phases at high temperature $T$ and zero baryon chemical potential $\mu_B$~\cite{Aoki:2006we}. Whereas at small $T$ and large $\mu_B$, a first-order phase transition is expected~\cite{Asakawa:1989bq}. Hence, there must be a critical point (CP) where the first-order phase transition line ends. Several ion-collision experimental programmes worldwide have been devoted to study the phase diagram at a wide range of $T$ and $\mu_B$ and to locate the CP. 
In particular, system size and energy scan programmes are ongoing at CERN SPS and BNL RHIC \cite{Adam:2020unf,Luo:2017faz, Davis:2019mlt}. 
One of the main goals of this scanning programs is to find evidence for 
large fluctuations near the CP in analogy to the phenomenon
of critical opalescence in the conventional matter \cite{PhysRevLett.19.555, Antoniou:2006zb}. 
In the future, the Nuclotron-based Ion Collider Facility (NICA) at JINR, Dubna and the Compressed Baryonic
Matter (CBM) at the Facility for Antiproton and Ion Research (FAIR) at GSI will join the study \cite{Ablyazimov:2017guv}.

Several experimental observables were proposed to search the CP, such as event-by-event fluctuations of conserved charges and power-law fluctuations within the framework of an intermittency method \cite{Bialas:1985jb, Bialas:1988wc, Satz:1989vj, Gupta:1990bi}. In the present work, we will focus only on the intermittency method. \revision{Particularly our interest is to show that the effect of momentum resolution is much stronger than intuitive expectations, see for example Ref.~\cite{EHSNA22:1993dgl}. Further, we study the dependence of the effect on two important parameters, the ratio of correlated to uncorrelated particles and the strength of the correlation given by the intermittency index.}

The paper is organized as follows. In Sec.~\ref{sec:intermittency}, we will briefly discuss the intermittency method. Modification of observable due to momentum resolution will be discussed in Sec.~\ref{sec:MomentumResolution} followed by the results in Sec.~\ref{sec:results}. Sec.~\ref{sec:conclusion} summarizes and concludes the paper.

\subsection{Intermittency to locate CP}
\label{sec:intermittency}
In this analysis method, observable of our interest is the scaled factorial moments.
The $q$-th order scaled factorial moment is defined as follows:
\begin{equation}
F_q (M) = \frac{ \left\langle \frac{1}{M^D} \sum_{i=1}^{M^D} n_i (n_i-1)...(n_i-q+1) \right\rangle }{ \left\langle \frac{1}{M^D} \sum_{i=1}^{M^D} n_i \right\rangle ^q },
\end{equation}
where $D$ is the dimension of the momentum space, $M^D$ is the number of bins in the momentum space, $n_i$ is the number of particles in $i$-th bin. Here $\langle...\rangle$ denote the average over events.
At the critical point, a power-law dependence of the scaled factorial moment on $M^D$ 
is expected:
\begin{equation}\label{eq:powerlaw}
F_q (M) = (M^{D})^{\phi_q}  ~,  
\end{equation}
when $M^D \gg 1$. The corresponding intermittency indices $\phi_q$ should follow the relation
\begin{equation}
 D \phi_q = (q-1) d_q ~,
\end{equation}
where $d_q$ is the anomalous fractal dimension \cite{DeWolf:1995nyp}.

The goal of the present work is to study the effect of momentum resolution on factorial moments. 
A Critical Monte Carlo Toy model is introduced. In the model, correlated pairs in each event are generated using the two particle distribution function:
\begin{equation}\label{eq:corr}
 \rho(X_1,X_2) =  \frac{\rho({X_1}) \rho({X_2})}{ \left| X_1-X_2 \right| ^{\phi_2} + \epsilon}~,
\end{equation}
where $\rho(X)$ is a single particle distribution in $X$.
Furthermore, we assume that
$X$ stands for dimensionless transverse momentum  $X \equiv p_T$/(1~GeV/$c$) and
it is distributed as
\begin{equation}\label{eq:rho}
 \rho(X) = C X e^{-6X},
\end{equation}
where $C$ is a normalization constant. The power $\phi_2$ in Eq.~\ref{eq:corr} is the intermittency index or the critical exponent. We have taken $\phi_2 = 0.2 - 0.8$ in the present analysis.
In the denominator of Eq.~\ref{eq:corr}, a small number $\epsilon = 10^{-6}$ is added to avoid singularity for $X_1 = X_2$. 
Note that the distribution of $X$ is highly non-uniform.

In the intermittency method, one-dimensional distribution of $X$ is divided into $M$ number of bins, and the second scaled factorial moment in one dimension is calculated using the definition 

\begin{equation}
F_2 (M) = \frac{ \left\langle \frac{1}{M} \sum_{i=1}^{M} n_i (n_i-1) \right\rangle }{ \left\langle \frac{1}{M} \sum_{i=1}^{M} n_i \right\rangle ^2 } .
\end{equation}
Equivalently one can write:
\begin{equation}\label{eq:F2}
F_2 (M) = \frac{2M}{ \langle N \rangle^2 } \langle N_{pp}(M) \rangle~,
\end{equation}
where $N$ and $N_{pp}(M)$ are multiplicity and the total number of particle pairs in M bins in an event, respectively.

\subsection{Momentum resolution}\label{sec:MomentumResolution}
In an experiment, true particle momentum is never measured. Instead, a momentum smeared by numerous stochastic effects related to the measurement process is extracted.
Here $X^s$ is the smeared momentum, and the difference between smeared and the true momenta is defined as 
\begin{equation}
 \Delta = X^{s} - X~.
\end{equation}
The $\Delta$ is not a constant but fluctuates in each measurement. We assume that $\Delta$ follows a Gaussian distribution with mean zero and standard deviation $\Sigma$:
\begin{equation}\label{eq:fd}
 f(\Delta) = \frac{1}{ \sqrt{2 \pi} \Sigma} e^{\frac{-\Delta^2}{2\Sigma^2}} ~.
\end{equation}
The dimensionless quantity $\Sigma$ is defined as $\Sigma \equiv \sigma$/(1~GeV/$c$) where $\sigma$ is the resolution of the measurement.

For correlated particles, the two-particle distribution function for true transverse momentum is given by Eq.~\ref{eq:corr}.
If particles are uncorrelated, then the two-particle distribution function is just the product of the individual distribution functions.
The two-particle distribution function for smeared transverse momentum distribution $\rho(X_1^s,X_2^s)$ will be different from $\rho(X_1,X_2)$ due to the resolution $\sigma$. Hence, $F_2$'s of $X$ and $X^s$ are expected to be different. Next, we will study this difference considering different values of $\sigma$. \revision{ The effect of detector resolution on factorial moments was discussed in Ref.~\cite{EHSNA22:1993dgl}. Transverse momentum was measured in the range  $-18<\ln (p_T/(1~GeV/c))^2< 6$ and the experimental resolution in 
$\ln (p_T/(1~GeV/c))^2$ was between 0.05 and 0.27. The maximum $M$ used in this work was 50. Therefore minimum bin width was 0.48. Since it was greater than the resolution, it was assumed that result would be unaffected by the resolution. Contrary to this assumption, in this work, we show that even if the bin width is much greater than the resolution, $F_2$ is significantly modified due to the effect of the resolution. In other words, the effect of the resolution is much stronger than the intuitive expectation discussed in Ref.~\cite{EHSNA22:1993dgl}.}

\section{Results}
\label{sec:results}

\begin{figure}[ptb]
\centering
\begin{center}
\includegraphics[width=0.48\textwidth]{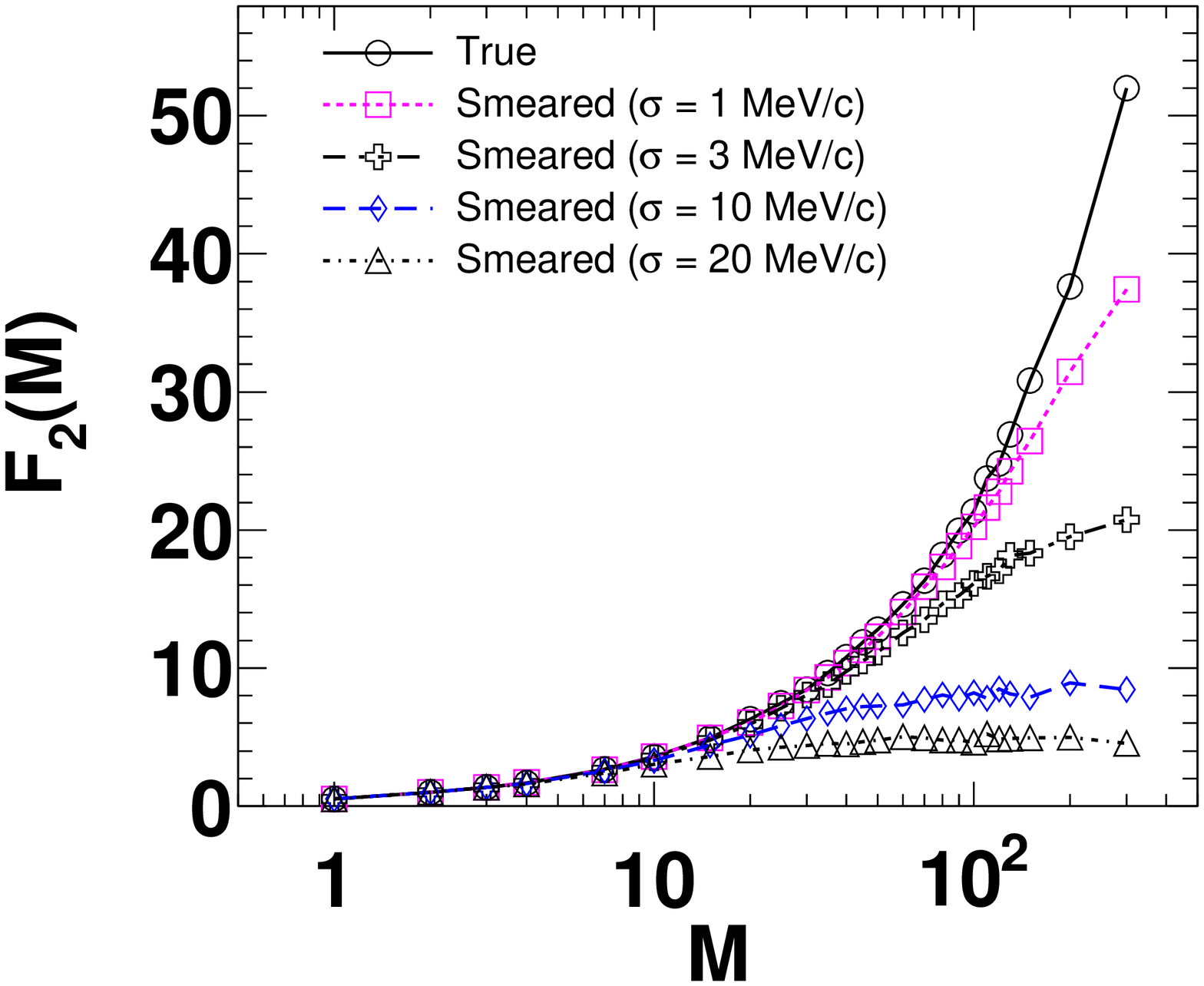}
\includegraphics[width=0.48\textwidth]{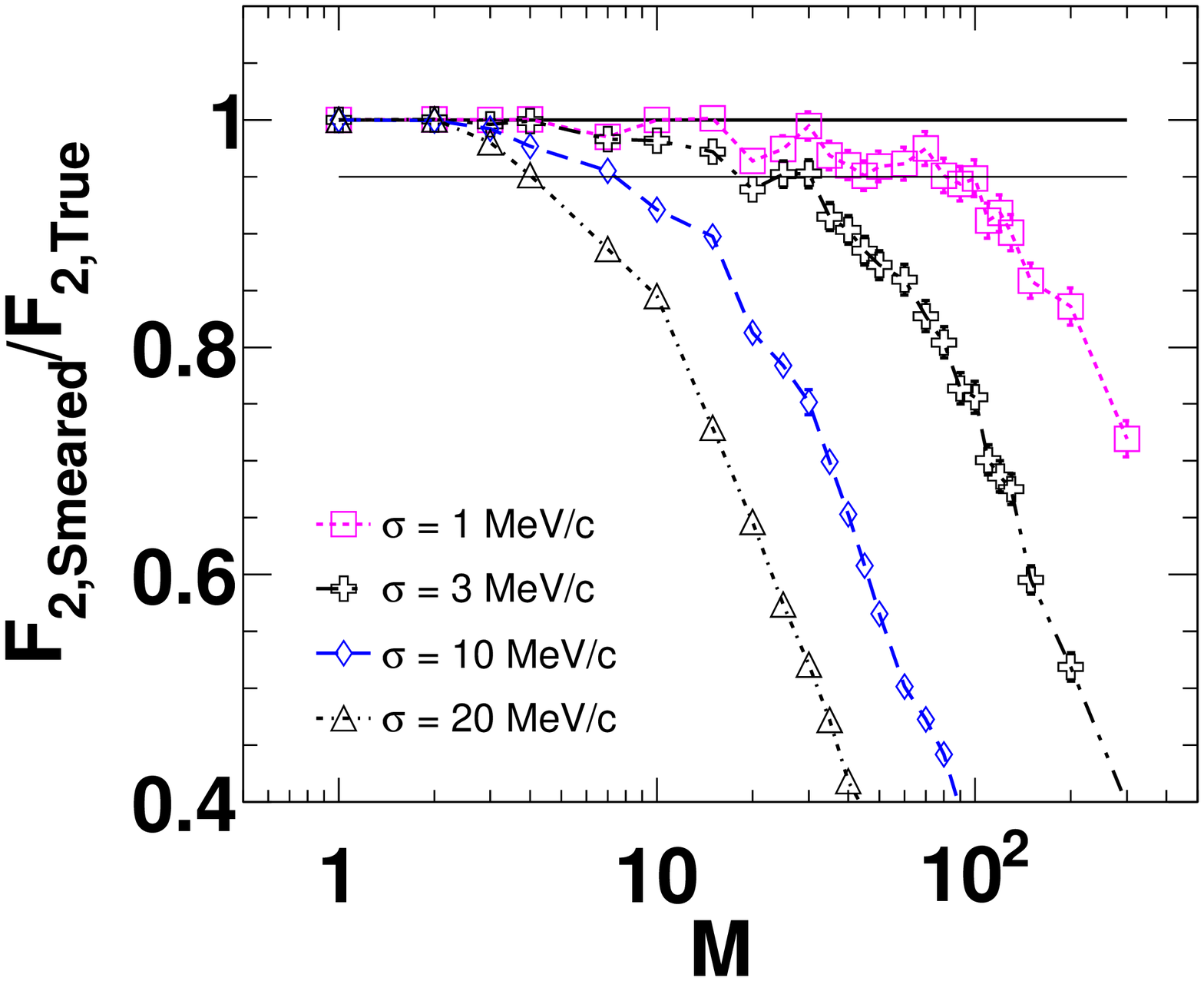}
 \end{center}
 \caption{Left panel shows $F_2$ vs $M$ for different values of standard deviation of momentum smearing. $F_2(M)$ is calculated in bins in transverse momentum $X$. Each event contains only two correlated particles ($N_c =2$). Right panel shows the dependence of ratios of smeared and true $F_2$ with $M$.}
\label{fig:F2_nonuniform}
\end{figure}

\subsection{Scenario 1: One correlated pair per event}


In numerical studies, we have varied $\sigma$ from 1~MeV/c up to~20 MeV/c. 
\revision{In an experiment, the detector resolution depends on several factors like detector topology and its resolution, the strength of the magnetic field and variables used. Typical transverse momentum resolution of the NA61/SHINE experiment at CERN SPS is of the order of $10$ MeV/c~\cite{Abgrall:2014xwa}.}
For the present study, $X$ is always generated within $0-1$ GeV/$c$.
The left panel of Fig.~\ref{fig:F2_nonuniform} shows the variation of $F_2$ with $M$. $F_2$ is calculated in bins of transverse momentum $X$. `True' corresponds to the result related to $X$, whereas `Smeared' represents the same for $X^s$. For all the figures shown in this paper, $F_2$ at each point is calculated from 10000 independent events. For this particular case, each event consists of one correlated pair, i.e., the number of correlated particles in each event is $N = N_c =2$.
Statistical error is calculated using the standard error propagation technique. For true $X$, at $M =1$, $F_2 = 0.5$ since $N = 2$ and $N_{pp} =1$ in each event. True $F_2$ increases rapidly with $M$ and shows a power-law behaviour at large $M$. \revision{Error bars are not visible in this plot because they are smaller than the symbol size.} At small $M$, $F_2$'s of true and smeared momenta are almost the same. As $M$ increases, smeared $F_2$ is suppressed compared to the true $F_2$ and the suppression increases with the increase of $\sigma$. To see the suppression clearly, we have shown the ratios of smeared and true $F_2$ in the right panel of Fig.~\ref{fig:F2_nonuniform}. A ratio is less than one indicates suppression. A line at 0.95 is drawn to quantify the deviation, which is the reference line of 5\% deviation. For $\sigma = 1$ MeV/$c$ more than 5\% deviation is observed when $M$ is larger than 100. As $\sigma$ increases, a similar deviation occurs at much smaller $M$. For example, for $\sigma = 20$ MeV/$c$, 5\% deviation is observed when $M$ is around 5.

\begin{figure}[ptb]
\centering
\begin{center}
\includegraphics[width=0.48\textwidth]{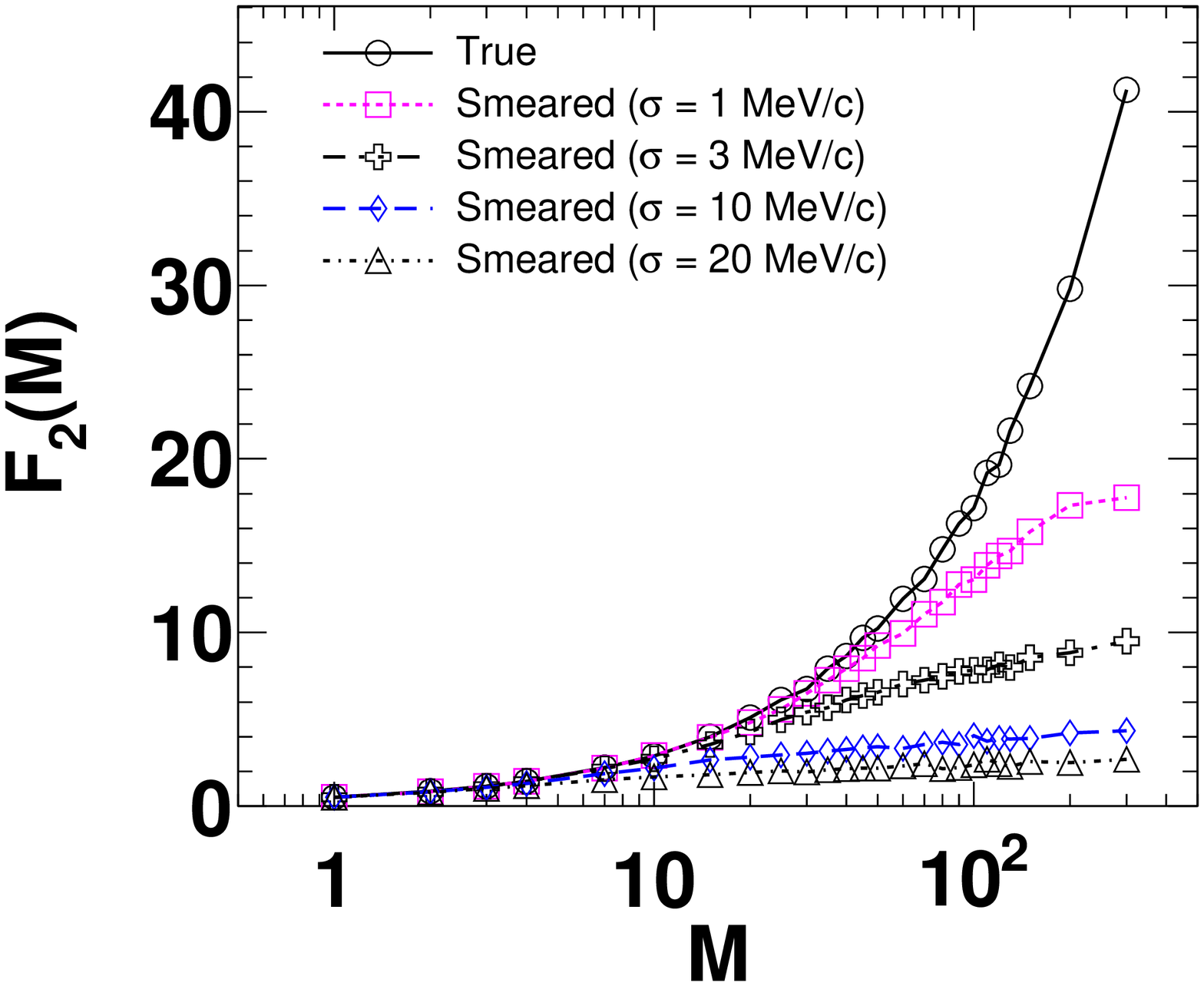}
\includegraphics[width=0.48\textwidth]{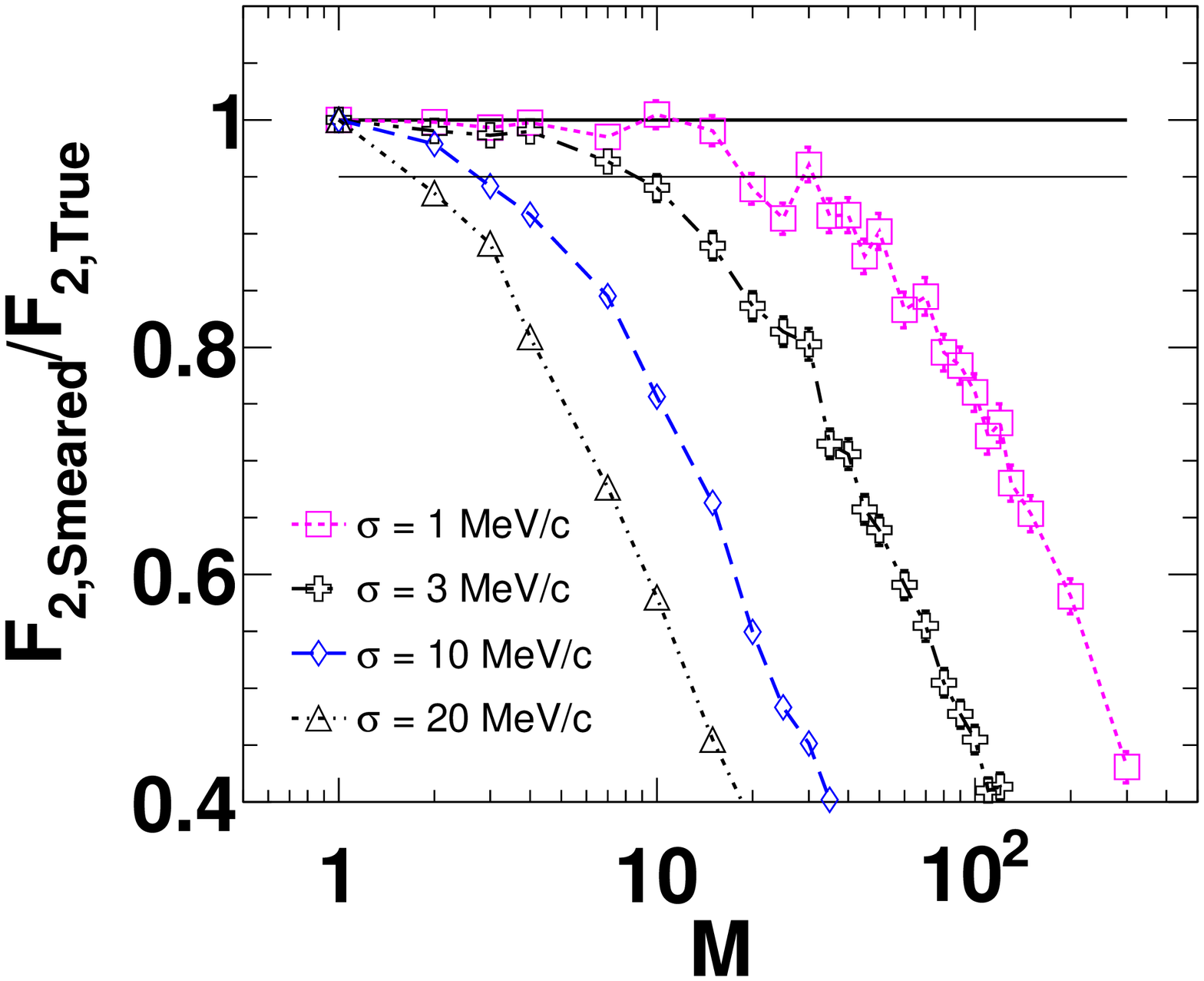}
 \end{center}
 \caption{Same as Fig.~\ref{fig:F2_nonuniform} but binning is done in cumulative transverse momentum $Q_{X}$.}
\label{fig:F2_uniform}
\end{figure}

\begin{figure}[ptb]
\centering
\begin{center}
\includegraphics[width=0.48\textwidth]{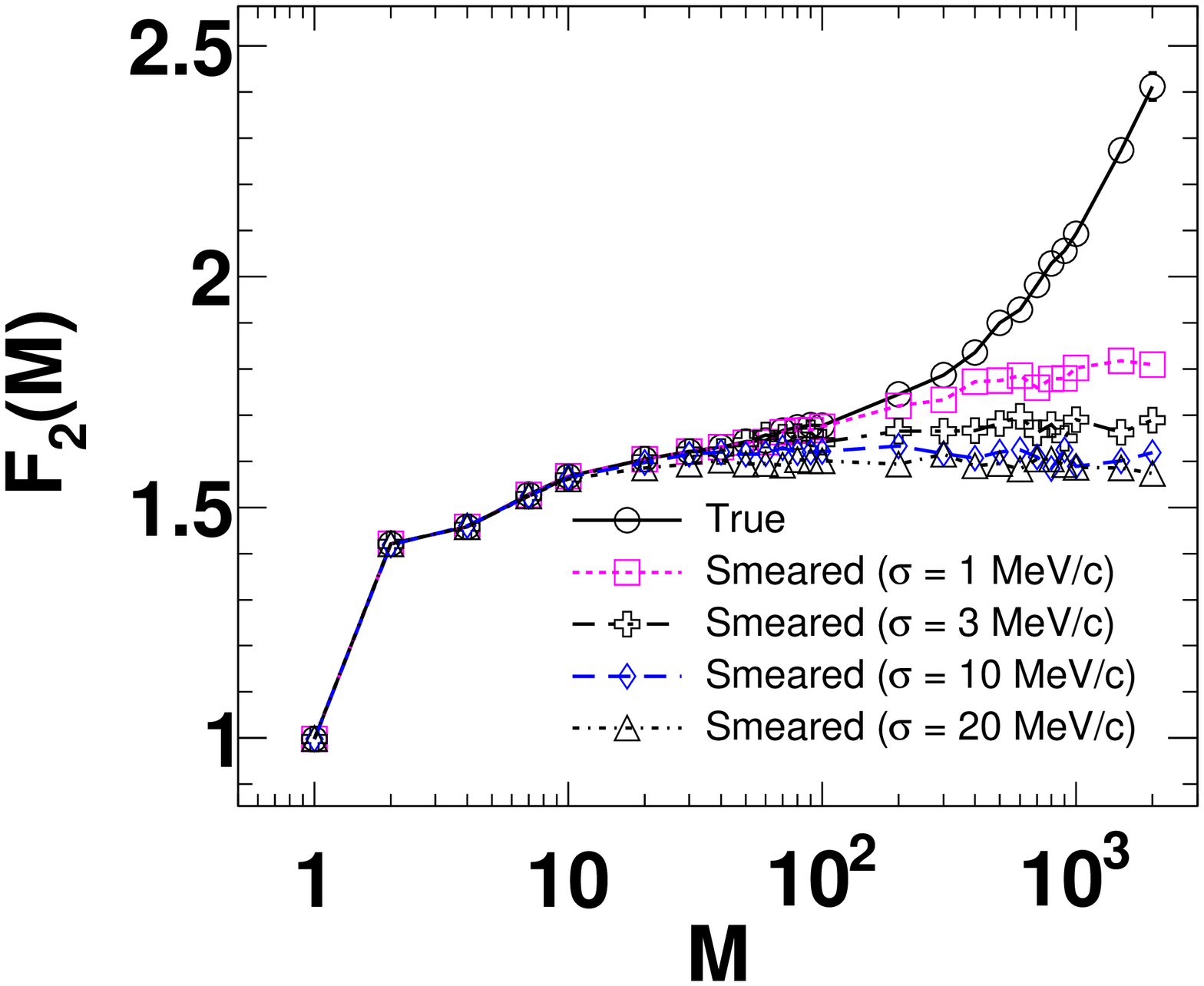}
\includegraphics[width=0.48\textwidth]{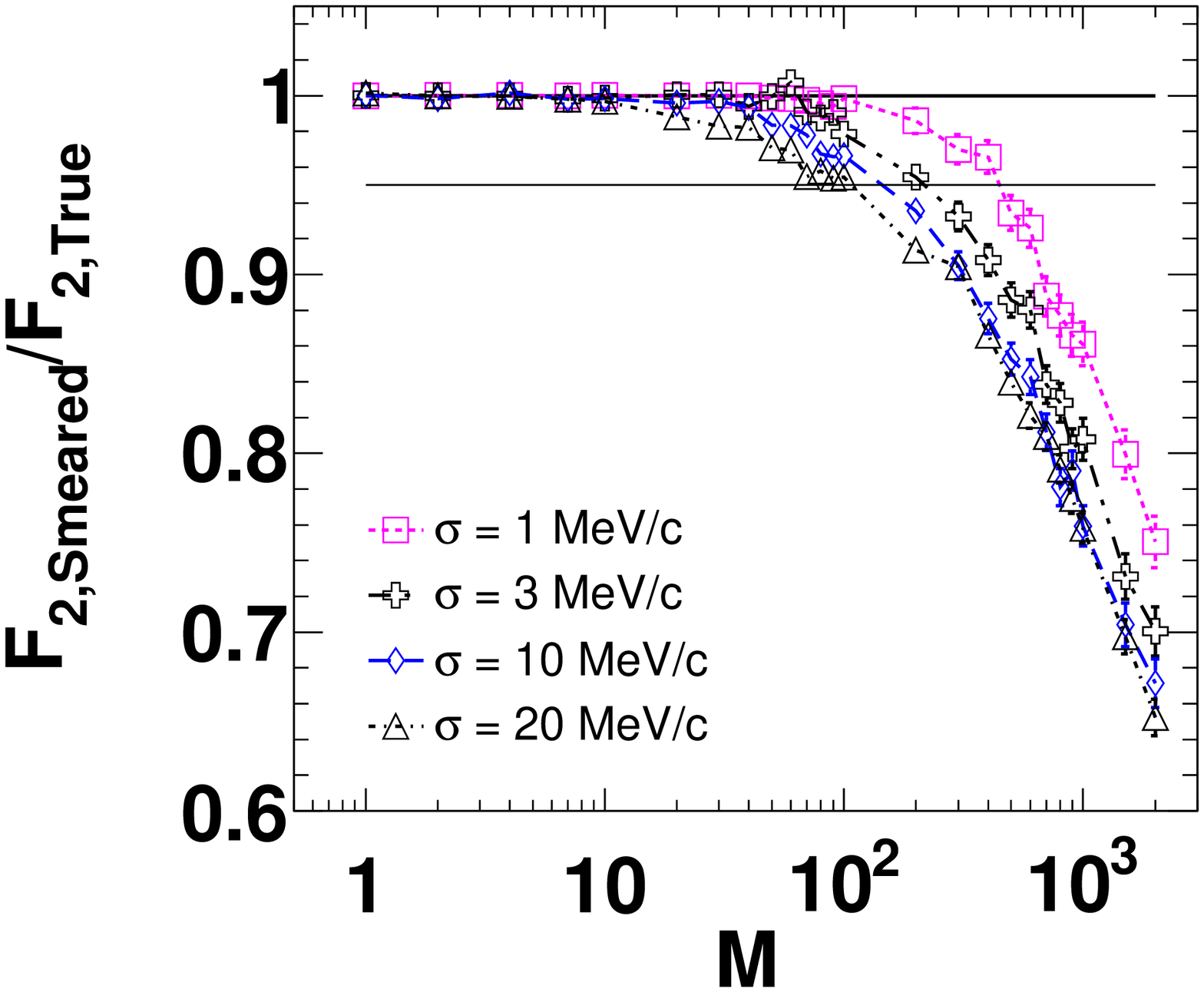}
 \end{center}
 \caption{Left panel shows $F_2$ vs $M$ for different values of standard deviation of momentum smearing where $F_2$ is calculated in bins in transverse momentum $X$. Each event contains both correlated ($N_c =2$) and uncorrelated ($\langle N_{uc}\rangle = 30$) particles. Right panel shows the dependence of ratios of corresponding smeared and true $F_2$ on $M$. 
 }
\label{fig:F2_nonuniform_with_background}
\end{figure}

\begin{figure}[ptb]
\centering
\begin{center}
\includegraphics[width=0.48\textwidth]{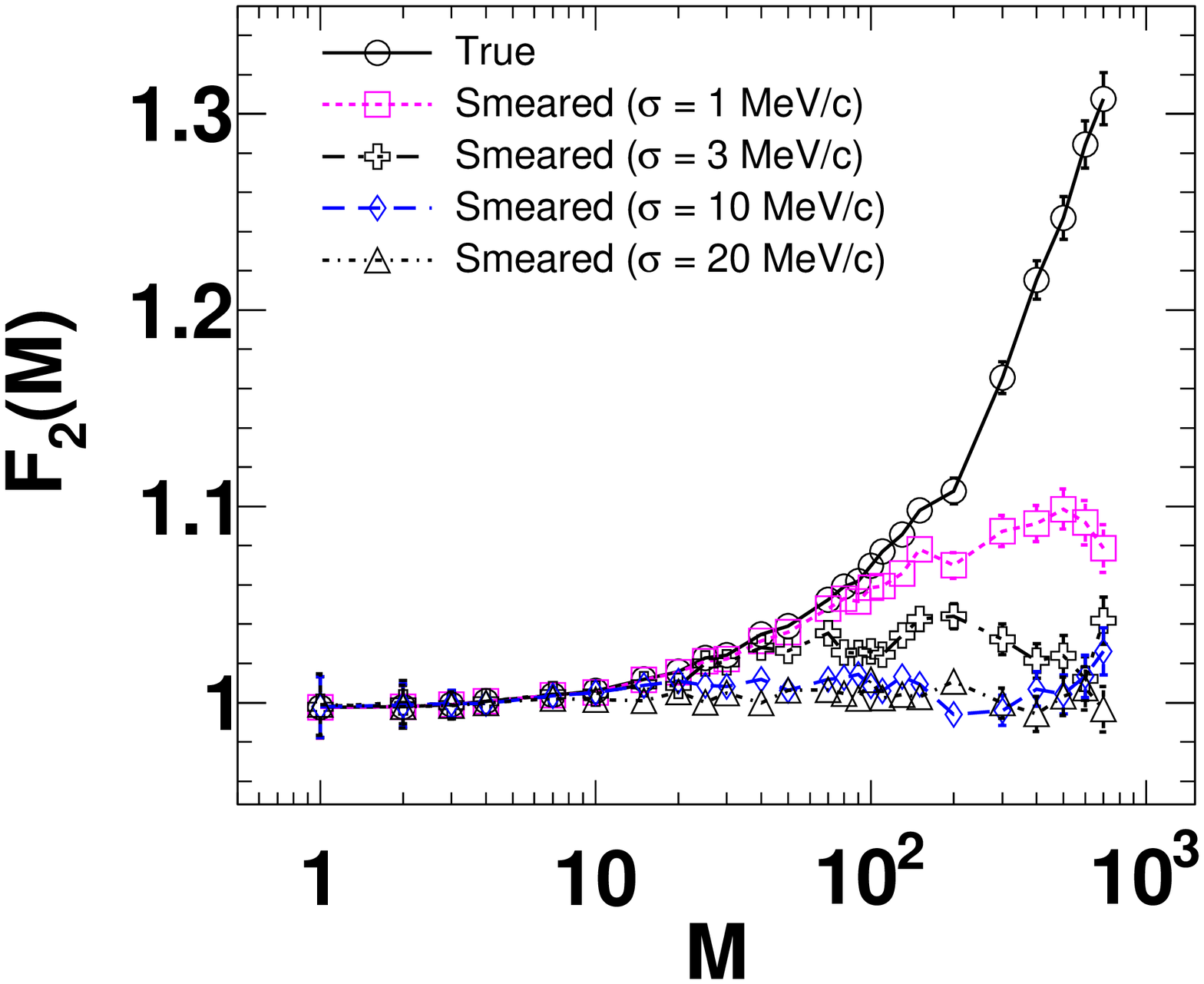}
\includegraphics[width=0.48\textwidth]{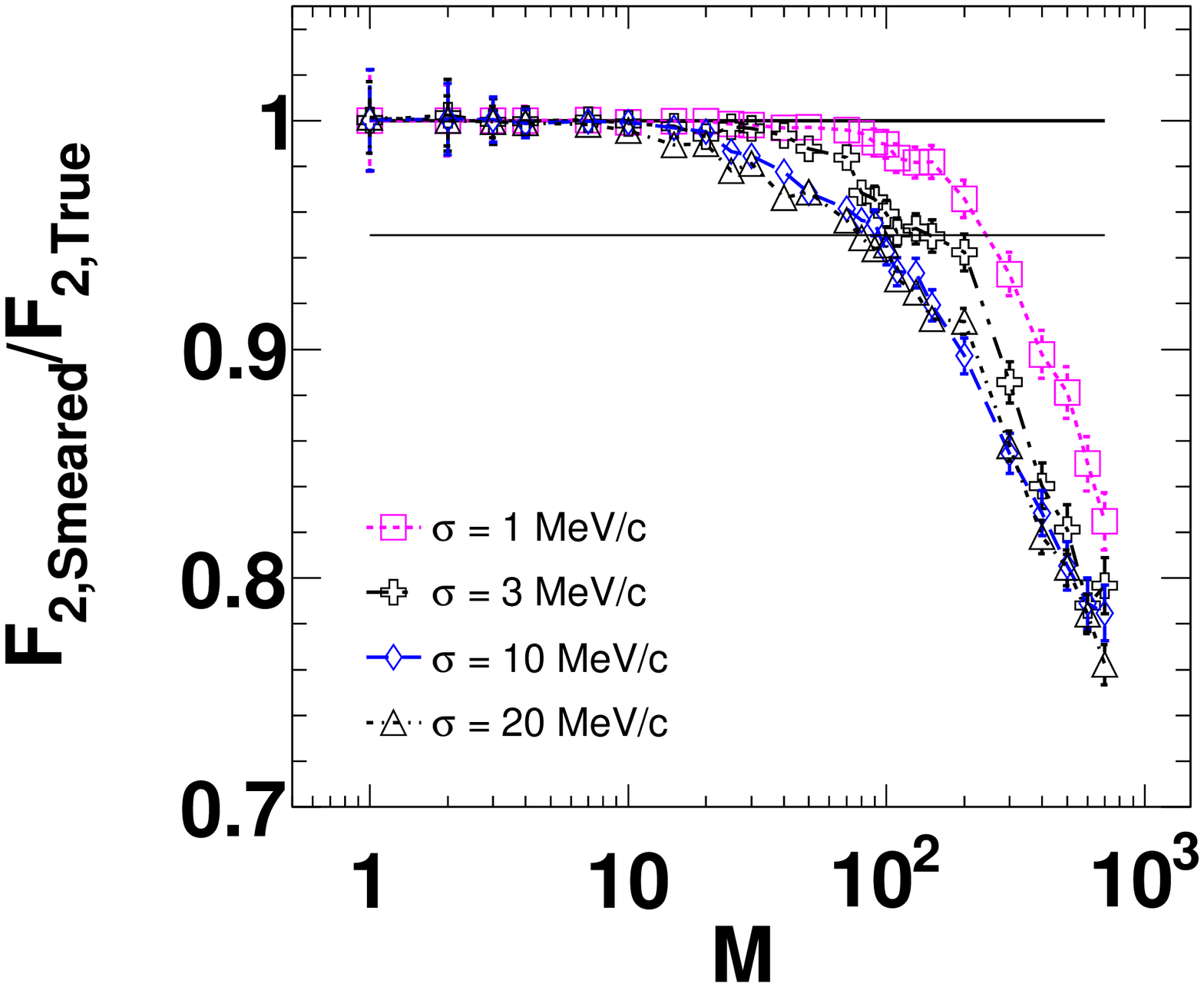}
 \end{center}
 \caption{Same as Fig.~\ref{fig:F2_nonuniform_with_background} but binning is done in cumulative transverse momentum $Q_X$.}
\label{fig:F2_uniform_with_background}
\end{figure}

Next, we have calculated $F_2$ for binning in cumulative transverse momentum distribution $Q_X$. A non-uniform transverse momentum distribution is converted into a uniform distribution using cumulative transformation \cite{Bialas:1990dk} defined as

\begin{equation}\label{eq:Q_X}
 Q_X = \frac{\int_{a}^{X} \rho(X) dX}{\int_{a}^{b} \rho(X) dX}~,
\end{equation}
where $a$ and $b$ are the lower and upper limits of $X$, respectively.
According to Eq.~\ref{eq:Q_X}, $Q_X$ varies between zero to one
and distribution of $Q_X$ becomes uniform.
The importance of the cumulative variable is that the scaling behaviour of $F_2$ with $M$ is obeyed, which is shown in our previous work \cite{Samanta:2021dxq}.

Now we have calculated $F_2$ binning particles in $Q_X$.
Results are presented in Fig.~\ref{fig:F2_uniform}. \revision{We observe that for $Q_X$, variation of true $F_2$ with $M$ follows power-law
function: 
\begin{equation}\label{eq:powerlaw-fit}
    f(M) = \alpha_{1} M^{\varphi_{2}} + \alpha_{2}
\end{equation}
where $\alpha_{1}, \alpha_{2}$ and $\varphi_{2}$ are parameters. The extracted intermittency index $\varphi_{2}$ $(\simeq 0.77)$ is close to the exponent 0.8, which was used to generate correlated pairs (Eq.~\ref{eq:corr}). The small difference is due to the small number $\epsilon$ present at the denominator of Eq.~\ref{eq:corr}. Smeared $F_2$ is suppressed compared to the true $F_2$, and the suppression increases with $M$. Suppression also increases with the increase of $\sigma$. Because of this suppression, power-law dependence is destroyed. If we still try to fit smeared $F_2$ using Eq.~\ref{eq:powerlaw-fit}, we would get the wrong value of the intermittency index. In the ratio plot, we observe that the effect of detector resolution is more pronounced in the case of $Q_X$ compared to Fig.~\ref{fig:F2_nonuniform}. The 5\% deviation between true and smeared $F_2$ occurs at even smaller $M$. For example, when $\sigma = 1$ MeV/c, the 5\% deviation is observed at $M > 30$ which was around 100 in case of non-uniform distribution.}

 Let us now discuss what we expect intuitively. \revision{For cumulative variable, we expect power-law (Eq. \ref{eq:powerlaw}) dependence of $F_2$ at any $M$ larger than one. But this is not true for the smeared $F_2$ due to the detector resolution effect.} When $M$ is 100, one bin corresponds to 10 MeV/$c$ as the $X$ is generated between 0-1 GeV/$c$. Hence when $\sigma \approx 10$ MeV/$c$, it is highly probable that we will not find two particles in the same bin. That means at $M \approx 100$, correlation is expected to be lost for $\sigma \approx 10$ MeV/$c$. However, we observe this deviation even at much smaller $M$, which is counter-intuitive. \revision{Definitely, one should not go beyond the limit when bin size is smaller than the $\sigma$. 
 However,
 even if $M$ is smaller than this limit, we should remember that smeared $F_2$ may be significantly smaller than the true one. As a result, from the smeared $F_2$ we would get an intermittency index much smaller than the actual value of $\phi_2$ in Eq.~\ref{eq:corr}. Therefore, to get the proper information of intermittency index, we have to first correct the momentum resolution effect. Once it is done, $F_2$ will show power-law behaviour at any $M$.}
  
\subsection{Scenario 2: Correlated and uncorrelated particles}
\revision{In heavy-ion collision, the created  system is expected to be inhomogeneous~\cite{Werner:2007bf, Aichelin:2008mi}. The freeze-out may be located at a distance from the CP. As a result, the CP may affect only a fraction of produced particles.}
\revision{ We will consider the above-mentioned scenario by assuming that both correlated and uncorrelated particles are present in an event. We will vary the fraction of correlated particles.}
In each event, some uncorrelated particles are added to one pair of correlated particles. The number of uncorrelated particles is not fixed but follows a Poisson distribution with mean at $\langle N_{uc}\rangle$. Left panel of Fig.~\ref{fig:F2_nonuniform_with_background} shows variation of $F_2$ with $M$ where $F_2$ is calculated in bins either in $X$ (True) or $X^s$ (Smeared). In this case, each event consists of $N_c =2$ and $\langle N_{uc}\rangle = 30$. Average number of particles in each event is $\langle N\rangle = 32$. $F_2$ at $M =1$ is close to unity. This is expected because for large $N$, number of pairs $N_{pp} = N(N-1)/2 \approx N^2/2$ and hence from Eq.~\ref{eq:F2} we can see that $F_2$ becomes $\approx 1$. With increase of $M$, $F_2$ of true $X$ increases. However, the magnitude is significantly smaller compared to Fig.~\ref{fig:F2_nonuniform}. \revision{Not only that, true $F_2$ shows power-law behaviour only when $M$ is large. In the region $M \lesssim 10$, behaviour of the curve is completely different compared to Fig.~\ref{fig:F2_nonuniform}.} Further, there is almost no effect of momentum resolution up to $M \approx 10$. For larger values of $M$, $F_2$ decreases with increase of $\sigma$. Ratio of true and measured $F_2$ is shown in the right panel of Fig.~\ref{fig:F2_nonuniform_with_background}. For $\sigma = 20$ MeV/$c$, deviation is more than 5\% as $M$ becomes larger than 500. Similar deviation is observed at around $M = 100$ for $\sigma = 1$ MeV/$c$. So here the 5\% deviation is occurring at larger $M$ compared to Fig.~\ref{fig:F2_nonuniform}.

Figure~\ref{fig:F2_uniform_with_background} is similar to Fig.~\ref{fig:F2_nonuniform_with_background}, but binning is done in $Q_X$. \revision{Variation of $F_2$ with $M$ shown in the left panel of Fig~\ref{fig:F2_uniform_with_background} is significantly different from that of Fig.~\ref{fig:F2_nonuniform_with_background}. Particularly, power-law (Eq. \ref{eq:powerlaw-fit}) type variation of true $F_2$ versus $M$ is now restored in all $M \ge 1$. Further, the extracted $\varphi_2 \simeq 0.73$, close to 0.8 used in the two-particle distribution function. For $M \sim 10$, smeared $F_2's$ are close to true $F_2$. However, at large $M$ they are significantly suppressed. The right panel of Fig.~\ref{fig:F2_uniform_with_background} shows the ratios of smeared and true $F_2$.}
Here the behaviour of curves is similar to Fig.~\ref{fig:F2_nonuniform_with_background}. However, the deviation occurs at slightly smaller $M$. A similar trend was also observed in Fig.~\ref{fig:F2_uniform} where we used only correlated particles.

\begin{figure}[ptb]
\centering
\begin{center}
\includegraphics[width=0.48\textwidth]{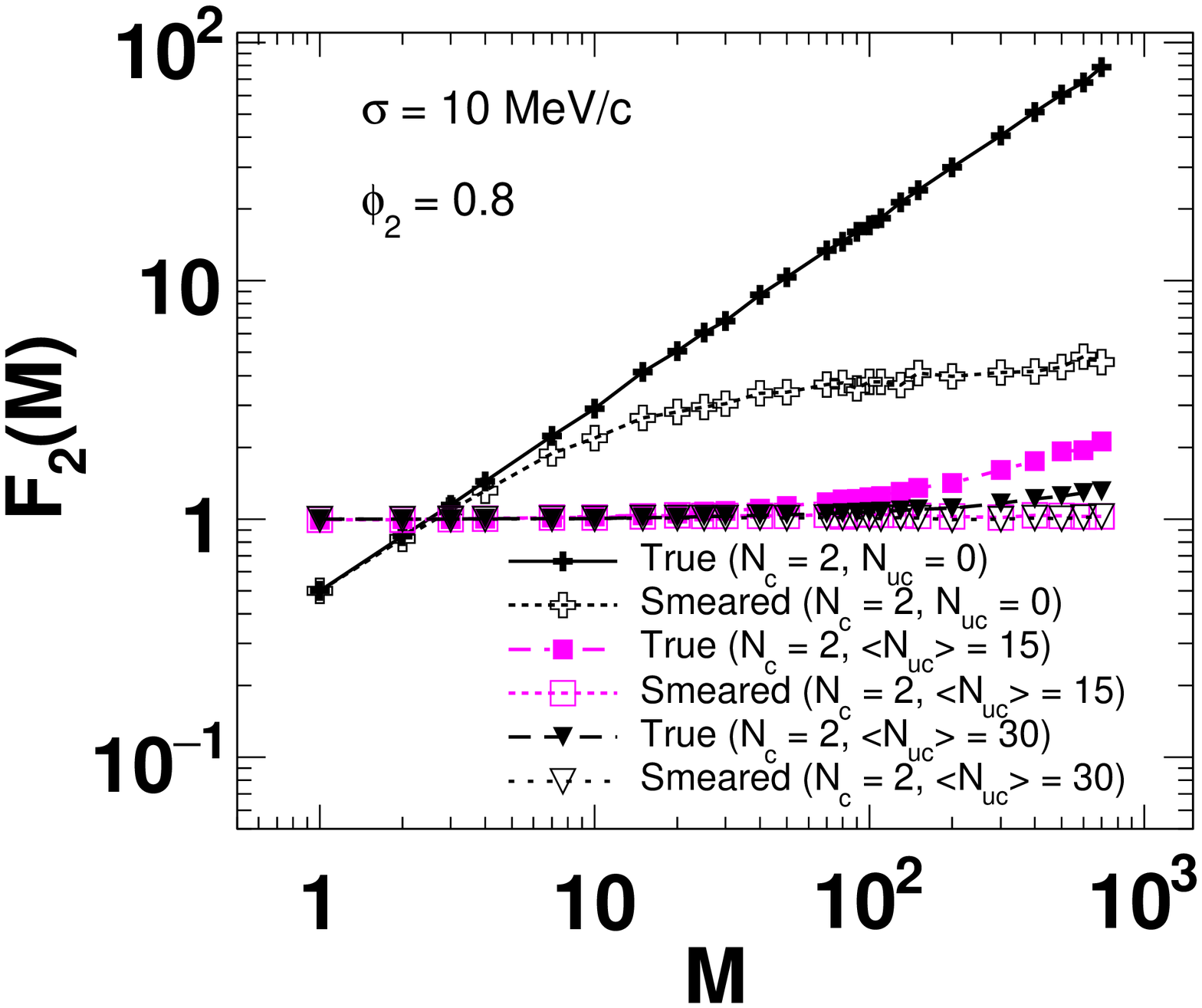}
\includegraphics[width=0.48\textwidth]{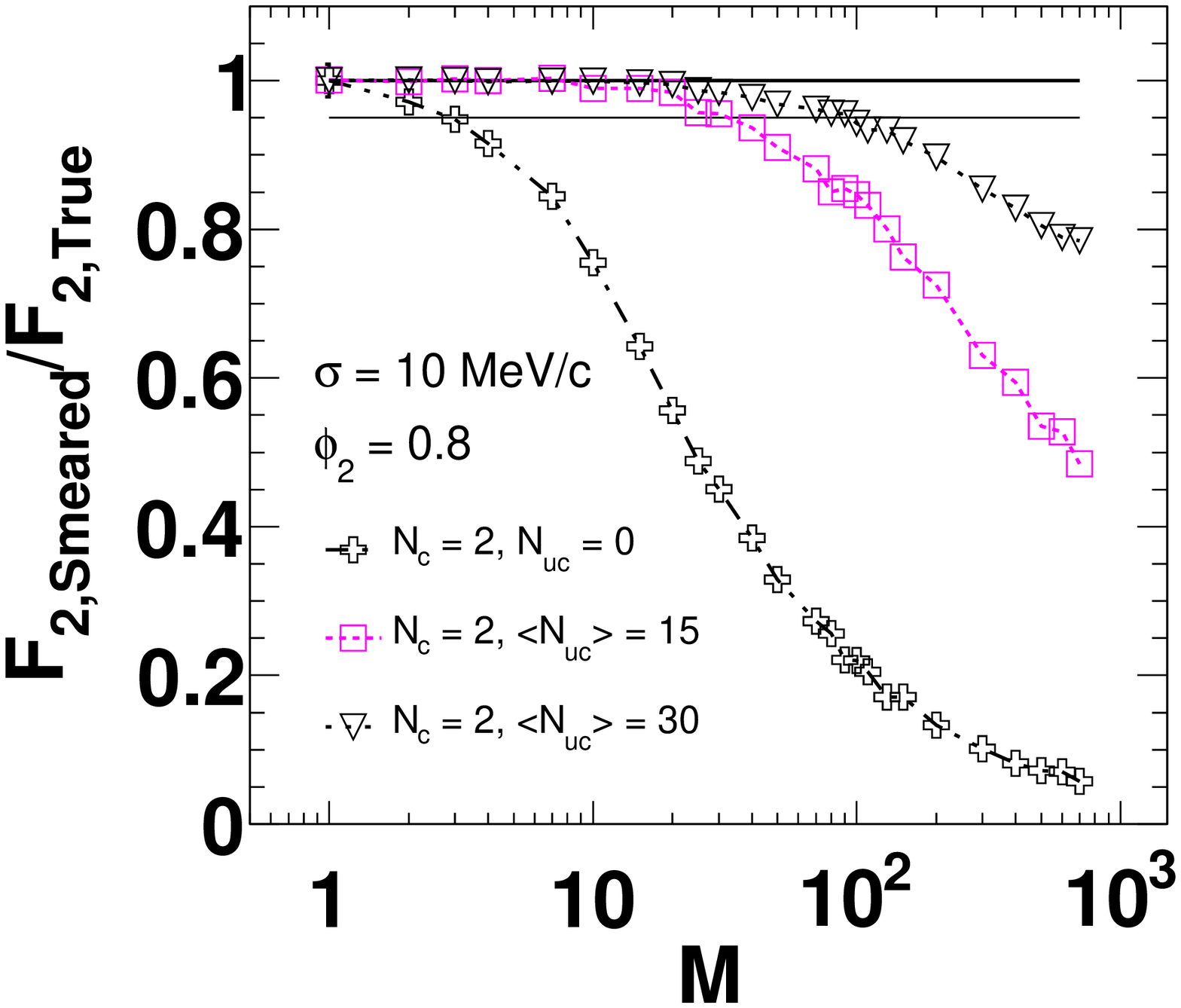}
 \end{center}
 \caption{Left panel shows $F_2$ vs $M$ for different values of mean number of uncorrelated particles $\langle N_{uc}\rangle$ added to one correlated pair of particles. The standard deviation of momentum smearing $\sigma =3$ MeV/$c$ for each smeared $F_2$. The right panel shows ratios of smeared to true $F_2$ with $M$.}
\label{fig:F2_uniform_with_different_background}
\end{figure}

We have analyzed a different number of uncorrelated particles $N_{uc}$ added to one pair of correlated particles $N_c$, as well. Results are shown in the left panel of Fig.~\ref{fig:F2_uniform_with_different_background}.
The cumulative variable is used here (the non-cumulative variable is not shown since, from previous Figs.~1-4, we already understand how the result is modified as we go from non-cumulative to the cumulative variable).
Three different cases are considered with $N_{uc} = 0$ (i.e, only correlated particles), $\langle N_{uc}\rangle = 15$ and $\langle N_{uc}\rangle = 30$. For the correlated pair, $\phi_2$ is taken as 0.8. For the first set, the total number of particles in each event is fixed and is equal to 2. Therefore, $F_2$ starts from 0.5. Already we have discussed that true $F_2(M)$ versus $M$ curve follows Eq.~\ref{eq:powerlaw-fit} when $N_{uc} = 0$. For the other two sets, the average number of particles in each event is 17 and 32, respectively. For these two cases, $F_2$ starts from one since the particle distribution is Poisson. We observe that true $F_2$ decreases substantially at large $M$ as we increase $N_{uc}$. \revision{Still in both the cases, true $F_2$ versus $M$ curves follow Eq.~\ref{eq:powerlaw-fit}. For $\langle N_{nc}\rangle = 15$ and 30 extracted intermittency indices are 0.78 and 0.73 respectively; both of which are close to 0.8. These values indicate that the intermittency index can be extracted from the true $F_2$ versus $M$ curve even in the presence of uncorrelated particles if the cumulative variable is used. This is one of the most important features of a cumulative variable. One could ask what would happen if we further increase $\langle N_{nc}\rangle$? In that case, true $F_2$ will be almost flat around 1; hence, it will be difficult to distinguish it from an uncorrelated system (for the fully uncorrelated system, $F_2 \approx 1$). For the smeared $F_2$, we have used $\sigma$ (=10 MeV/$c$) for all three sets. The smeared $F_2$ is suppressed compared to true $F_2$ in all three cases. The right panel of Fig.~\ref{fig:F2_uniform_with_different_background} shows the variation of the ratio of true to smeared $F_2$ with $M$ for a different number of uncorrelated particles keeping $\sigma$ (=10 MeV/$c$) fixed. At low $M$, the ratio is approximately one for all three cases. Ratio decreases with the increase of $M$. However, the decrease is happening relatively slowly as we increase $N_{uc}$. For the first set ($N_{uc} =0$), the ratio becomes less than 0.95 when $M$ greater than $4$. In case of the second set, where $\langle N_{uc}\rangle = 15$, the ratio is almost 1 up to $M \sim 7$. Then the ratio starts decreasing with a further increase of $M$, and it becomes less than 0.95 when $M$ is $\gtrsim 50$. The decrease of ratio is even slighter for the third set where $\langle N_{uc}\rangle = 30$. In this set, more than 5\% deviation is happening when $M > 110$. This figure clearly shows that the 5\% deviation shifts towards the higher values of $M$ as the ratio $N_c/N_{uc}$ is decreased.}

\subsection{Scenario 3: Only correlated particles with different $\phi_2$ }

\begin{figure}[ptb]
\centering
\begin{center}
\includegraphics[width=0.48\textwidth]{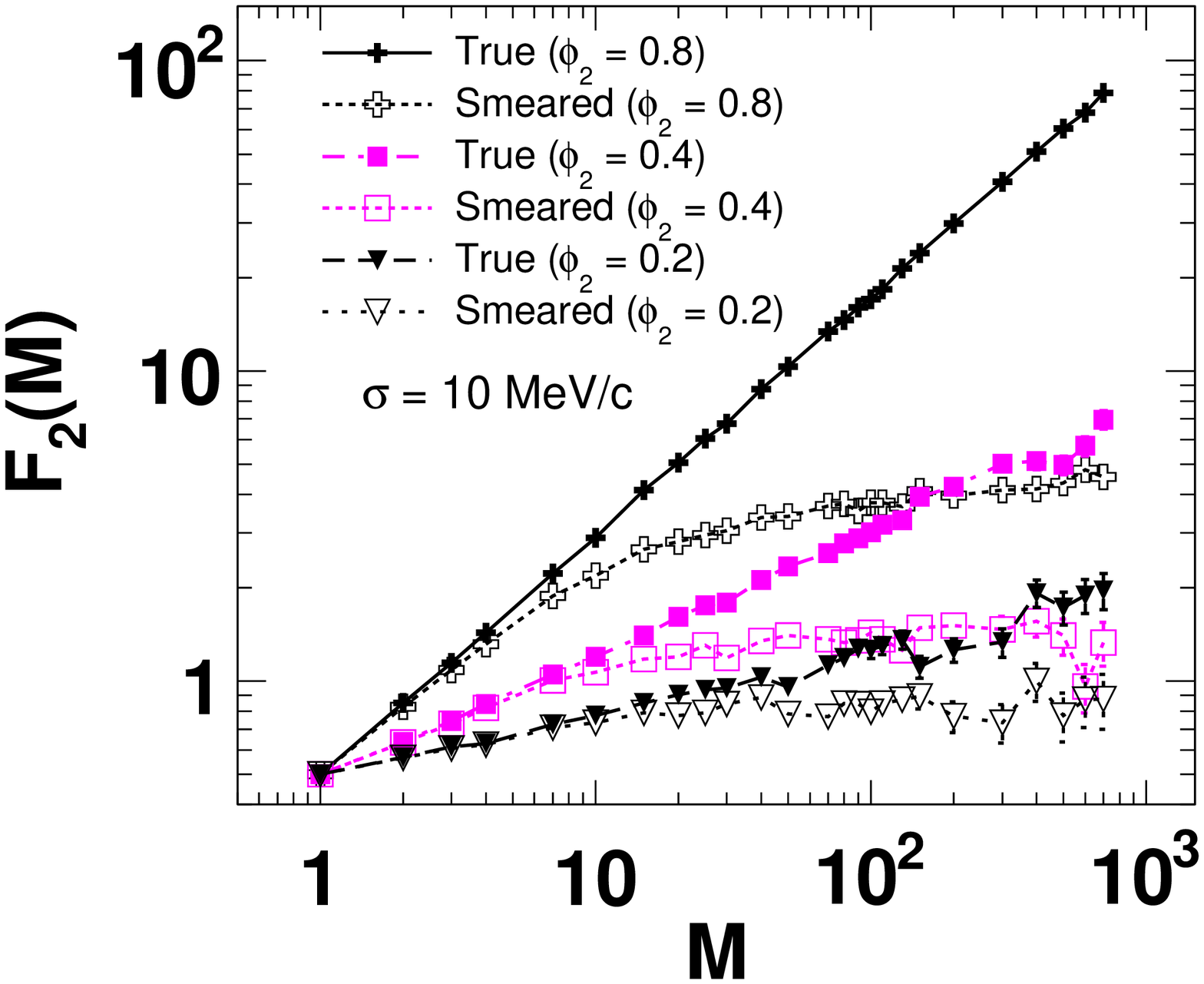}
\includegraphics[width=0.48\textwidth]{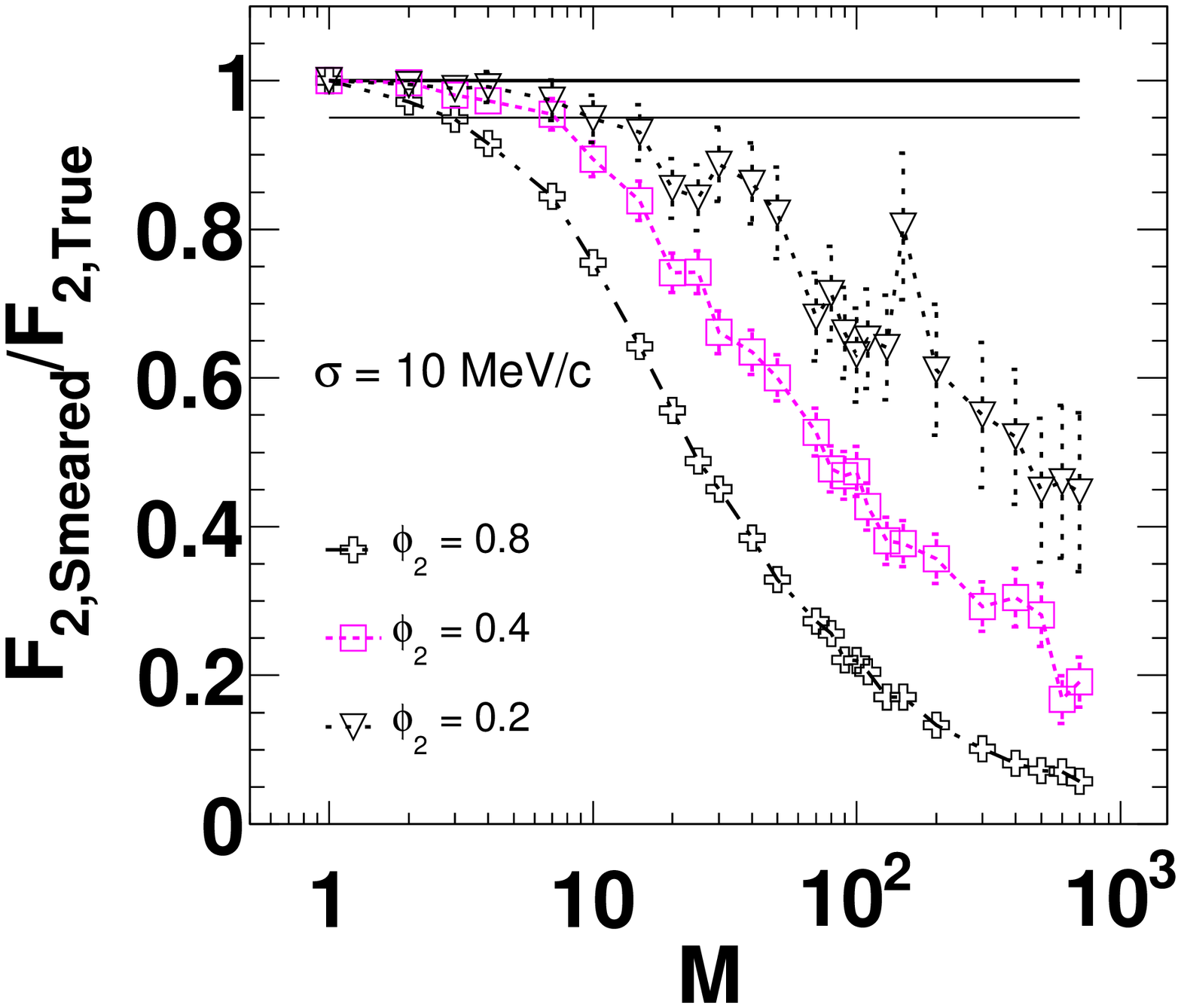}
 \end{center}
 \caption{Left panel shows variation of $F_2$ with $M$ for different values of $\phi_2$. Binning is done in $Q_X$. Each event contains only one pair of correlated particles. Fixed $\sigma = 10$ MeV/c is used for smeared case. Right panel shows variation of ratios of smeared to true $F_2$ with $M$.}
\label{fig:F2_uniform_with_different_phi}
\end{figure}

\revision{In this present work, correlated pairs generated by Eq.~\ref{eq:corr}. In this equation, correlation is controlled by the critical exponent $\phi_2$.  When $\phi_2 \to 0$, $\rho(X_1, X_2) \to \rho(X_1) \rho(X_2)$. This implies that the particles become uncorrelated. In other words, the effect of correlation decreases with a decrease of $\phi_2$, and in the limit $\phi_2 \to 0$ particles are uncorrelated. In this subsection, we will study the effect 
of resolution for different values of $\phi_2$.}

\revision{In the left panel of Fig.~\ref{fig:F2_uniform_with_different_phi} we show $F_2$ vs $M$ for three different sets with $\phi_2 = 0.8$, 0.4 and 0.2. $F_2$ is calculated in bins of $Q_X$. Here each event has only one pair of correlated particles and no uncorrelated particles are there. As a result $F_2 = 0.5$ at $M =1$ for all the three sets. True $F_2$ increases with increase of $M$ for all the three sets. However, at a particular $M$, true $F_2$ decrease with decrease of $\phi_2$. This behaviour is expected since Eq.~\ref{eq:powerlaw} says, $F_2$ becomes independent of $M$ at $\phi_2 =0$. Eq.~\ref{eq:powerlaw} also implies that the variation of true $F_2$ with $M$ should be a straight line in $\log-\log$ scale and indeed it is observed for all the three sets with three different $\phi_2$. This linear variation indicates that the power-law behaviour is there in all the three cases (we have already discussed this point for $\phi_2 = 0.8$). Further, the slope of the curve decreases with decrease of $\phi_2$. This is also expected, because in $\log-\log$ plot $\phi_2$ mimics the slope of the curve. With decreasing $\phi_2$ slope should decrease as indicated by Eq.~\ref{eq:powerlaw}. For smearing we use a constant resolution ($\sigma = 10$ MeV/$c$). As usual smeared $F_2$ is suppressed compared to true $F_2$ for all the three sets. The right panel of Fig.~\ref{fig:F2_uniform_with_different_phi} shows the ratio of true to smeared $F_2$ for three different $\phi_2$ values keeping $\sigma$ fixed. For $\phi_2 = 0.8$, the deviation of smeared $F_2$ is more than 5\% compared to true $F_2$ when $M$ is approximately 4 or more. The deviation is more than 5\% when $M \ge 10$ in case of $\phi_2 = 0.4$. In the third set, where $\phi_2 = 0.2$, the deviation is more than 5\% when $M \ge 20$. Therefore, as $\phi_2$ decreases, 5\% deviation shifts towards higher values of $M$.}

\revision{Let us now discuss the importance of the last two figures, Fig.~\ref{fig:F2_uniform_with_different_background} and Fig.~\ref{fig:F2_uniform_with_different_phi}. In the intermittency analysis, experimental data is usually configured with the model by varying $N_c/N_{uc}$ and $\phi_2$. In Figs.~\ref{fig:F2_uniform_with_different_background} and \ref{fig:F2_uniform_with_different_phi} we have done the same thing. We observe some
similarities in the last two figures \ref{fig:F2_uniform_with_different_background} and \ref{fig:F2_uniform_with_different_phi}. The right panel of Fig.~\ref{fig:F2_uniform_with_different_background} shows that 5\% deviation between smeared and true $F_2$ occurs at larger $M$ as we decrease $N_c/N_{uc}$ keeping $\phi_2$ fixed. (Note $\sigma$ is also fixed.) Similar trend is observed in the right panel of Fig.~\ref{fig:F2_uniform_with_different_phi} where we decrease $\phi_2$ keeping the number of particles fixed ($N_c = 2, N_{uc} =0$, so $N_{uc}/N_{c} =0$). That means decreasing $N_c/N_{uc}$ is equivalent to decreasing $\phi_2$ as far as ratio of smeared to true $F_2$ is concerned. In two extreme limits (1) $N_c/N_{uc} \to 0$ or (2) $\phi_2 \to 0$,  ratio of smeared and true $F_2$ will be one even at large $M$. In the first limiting case, true $F_2$ went towards one and smeared $F_2$ will also be one. So both true and smeared $F_2$ behave like an uncorrelated system in which $F_2$ is one when there are many particles in each event. On the other hand, in the second limiting case, when $\phi_2 \to 0$, the slope of true $F_2$ versus $M$ curve becomes 0. In the particular example of Fig.~\ref{fig:F2_uniform_with_different_phi} where we consider only one pair of correlated particles, true $F_2$ goes towards 0.5 in all $M$ as $\phi_2 \to 0$. Smeared $F_2$ will also be 0.5, and hence the ratio of smeared and true $F_2$ will be one. Note that the ratio of smeared to true $F_2$ is also one at small $M$, because of a completely different reason. For small $M$, smeared $F_2$ is close to true $F_2$ and both the curves show power-law behaviour in that region. Figures~\ref{fig:F2_uniform_with_different_background} and \ref{fig:F2_uniform_with_different_phi} will be useful to properly configure parameters of the intermittency method used in heavy-ion collision experiments.
}

\section{Summary and conclusion}
\label{sec:conclusion}

We have studied the effect of momentum resolution on the second scaled factorial moment. The study is done for both non-uniform transverse momentum distribution and its corresponding uniform cumulative distribution. 
We observed that smeared $F_2$ is significantly different from true $F_2$ for the momentum resolution considered in work. Smeared $F_2$ decreases with the increase of momentum resolution $\sigma$. To quantify the deviation of measured $F_2$ from true $F_2$, we have shown the ratios as well. The 5\% deviation of smeared $F_2$ from true $F_2$ is observed at smaller values of $M$ for uniform distribution than non-uniform distribution. The effect is stronger than our intuitive expectations.
Further, deviation shifts towards larger values of $M$ when the ratio $N_c/N_{uc}$ is decreased. A similar effect is observed when we decrease $\phi_2$ keeping other parameters fixed. The results allow to properly configure parameters of the intermittency method used for search for the critical point in heavy-ion collisions. \revision{Correcting the effect of momentum resolution from $F_2$ is not an easy task. The unfolding technique might be used for this purpose. 
}

\begin{acknowledgments} 
This work was supported by the Polish National Science Centre grant number 2018/30/A/ST2/00226.
S. S. is supported by the Polish National Agency for Academic Exchange through Ulam Scholarship with Agreement
No: PPN/ULM/2019/1/00093/U/00001.
\end{acknowledgments}

\bibliographystyle{ieeetr}
\bibliography{references}

\begin{thebibliography}{10}

\bibitem{Aoki:2006we}
Y.~Aoki, G.~Endrodi, Z.~Fodor, S.~Katz, and K.~Szabo, ``{The Order of the
  quantum chromodynamics transition predicted by the standard model of particle
  physics},'' {\em Nature}, vol.~443, pp.~675--678, 2006.

\bibitem{Asakawa:1989bq}
M.~Asakawa and K.~Yazaki, ``{Chiral Restoration at Finite Density and
  Temperature},'' {\em Nucl. Phys. A}, vol.~504, pp.~668--684, 1989.

\bibitem{Adam:2020unf}
J.~Adam {\em et~al.}, ``{Nonmonotonic Energy Dependence of Net-Proton Number
  Fluctuations},'' {\em Phys. Rev. Lett.}, vol.~126, no.~9, p.~092301, 2021.

\bibitem{Luo:2017faz}
X.~Luo and N.~Xu, ``{Search for the QCD Critical Point with Fluctuations of
  Conserved Quantities in Relativistic Heavy-Ion Collisions at RHIC : An
  Overview},'' {\em Nucl. Sci. Tech.}, vol.~28, no.~8, p.~112, 2017.

\bibitem{Davis:2019mlt}
N.~Davis, N.~Antoniou, and F.~Diakonos, ``{Recent Results from Proton
  Intermittency Analysis in Nucleus--Nucleus Collisions from NA61 at CERN
  SPS},'' {\em Acta Phys. Polon. B}, vol.~50, pp.~1029--1040, 2019.

\bibitem{PhysRevLett.19.555}
H.~Brumberger, N.~G. Alexandropoulos, and W.~Claffey, ``Critical opalescence of
  liquid sodium-lithium mixtures,'' {\em Phys. Rev. Lett.}, vol.~19,
  pp.~555--556, Sep 1967.

\bibitem{Antoniou:2006zb}
N.~Antoniou, F.~Diakonos, A.~Kapoyannis, and K.~Kousouris, ``{Critical
  opalescence in baryonic QCD matter},'' {\em Phys. Rev. Lett.}, vol.~97,
  p.~032002, 2006.

\bibitem{Ablyazimov:2017guv}
T.~Ablyazimov {\em et~al.}, ``{Challenges in QCD matter physics --The
  scientific programme of the Compressed Baryonic Matter experiment at FAIR},''
  {\em Eur. Phys. J. A}, vol.~53, no.~3, p.~60, 2017.

\bibitem{Bialas:1985jb}
A.~Bialas and R.~B. Peschanski, ``{Moments of Rapidity Distributions as a
  Measure of Short Range Fluctuations in High-Energy Collisions},'' {\em Nucl.
  Phys. B}, vol.~273, pp.~703--718, 1986.

\bibitem{Bialas:1988wc}
A.~Bialas and R.~B. Peschanski, ``{Intermittency in Multiparticle Production at
  High-Energy},'' {\em Nucl. Phys. B}, vol.~308, pp.~857--867, 1988.

\bibitem{Satz:1989vj}
H.~Satz, ``{Intermittency and Critical Behavior},'' {\em Nucl. Phys. B},
  vol.~326, pp.~613--618, 1989.

\bibitem{Gupta:1990bi}
S.~Gupta, P.~La~Cock, and H.~Satz, ``{The Search for intermittency in the
  finite size Ising model},'' {\em Nucl. Phys. B}, vol.~362, pp.~583--598,
  1991.

\bibitem{EHSNA22:1993dgl}
N.~M. Agababyan {\em et~al.}, ``{Factorial moments, cumulants and correlation
  integrals in pi+ p and K+ p interactions at 250-GeV/c.},'' {\em Z. Phys. C},
  vol.~59, pp.~405--426, 1993.

\bibitem{DeWolf:1995nyp}
E.~De~Wolf, I.~Dremin, and W.~Kittel, ``{Scaling laws for density correlations
  and fluctuations in multiparticle dynamics},'' {\em Phys. Rept.}, vol.~270,
  pp.~1--141, 1996.

\bibitem{Abgrall:2014xwa}
N.~Abgrall {\em et~al.}, ``{NA61/SHINE facility at the CERN SPS: beams and
  detector system},'' {\em JINST}, vol.~9, p.~P06005, 2014.

\bibitem{Bialas:1990dk}
A.~Bialas and M.~Gazdzicki, ``{A New variable to study intermittency},'' {\em
  Phys. Lett. B}, vol.~252, pp.~483--486, 1990.

\bibitem{Samanta:2021dxq}
S.~Samanta, T.~Czopowicz, and M.~Gazdzicki, ``{Scaling of factorial moments in
  cumulative variables},'' {\em Nucl. Phys. A}, vol.~1015, p.~122299, 2021.

\bibitem{Werner:2007bf}
K.~Werner, ``{Core-corona separation in ultra-relativistic heavy ion
  collisions},'' {\em Phys. Rev. Lett.}, vol.~98, p.~152301, 2007.

\bibitem{Aichelin:2008mi}
J.~Aichelin and K.~Werner, ``{Centrality Dependence of Strangeness Enhancement
  in Ultrarelativistic Heavy Ion Collisions: A Core-Corona Effect},'' {\em
  Phys. Rev. C}, vol.~79, p.~064907, 2009.
\newblock [Erratum: Phys.Rev.C 81, 029902 (2010)].

\end{thebibliography}
\end{document}